\documentstyle[12pt,epsf]{article}
\if@twoside
 \else
 \fi
 
 \parskip \medskipamount
\newcommand{\be}{\begin{eqnarray}}
\newcommand{\ee}{\end{eqnarray}}

\begin{document}

\hbox{} \nopagebreak
\vspace{-3cm} \addtolength{\baselineskip}{.8mm} \baselineskip=24pt 
\begin{flushright}
{\sc OUTP-98 59P} \\
\end{flushright}

\vspace{40pt}
\begin{center}
{\large {\sc {\bf Strings and string breaking in  2+1
dimensional nonabelian theories.}}} \baselineskip=12pt \vspace{34pt}

A. Kovner$^1$ and B. Rosenstein$^2$ \vspace{24pt}

$^{1}$Theoretical Physics, Oxford University, 1 Keble Road, Oxford, OX1 3NP,
UK\\[0pt]
$^{2}$ Theoretical Physics Center and Electrophysics Department, National
Chiao Tung University, Hsinchu, Taiwan 30043, ROC\\[0pt]
\vspace{60pt}
\end{center}

\vspace{40pt}
\begin{abstract}
We consider properties of confining strings in 2+1 dimensional $SU(2)$
nonabelian gauge theory with the Higgs field in adjoint representation. The
analysis is carried out in the context of effective dual Lagrangian which
describes the dynamics of t'Hooft's $Z_{N}$ vorices. We point out that the
same Lagrangian should be interpreted as an effective Lagrangian for the
lightest glueballs. It is shown how the string tension for a fundamental
string arises in this description. We discuss the properties of the adjoint
string and explain how its breaking occurs when the distance between the
charges exceeds a critical value. The interaction between the fundamental
strings is studied. It is shown that they repell each other in the weak coupling regime. We
argue that in the confining regime (pure Yang-Mills theory, or a theory
with a heavy Higgs field) the strings actually attract each other and the
crossover between the two regimes corresponds to the crossover between the
dual superconductors of first and second kind.
\end{abstract}

\vfill

\newpage

\section{Introduction.}

Confinement is certainly one of the most striking qualitative features of
nonabelian gauge theories. Still a reasonable understanding of this
phenomenon is sorely missing. Although several possible mechanisms have
been discussed over the years, we do not have a good qualitative
understanding derived from a nonabelian gauge theory itself. The dual
superconductivity framework \cite{thooft1},\cite{mandelstam} is currently the
most popular way to view the phenomenon of confinement. Lattice studies of
magnetic monopoles in recent years seem to point to relevance of the
monopole condensation to confining properties of nonabelian gauge theories
\cite{lattice}. The exact solution of $N=2$ SUSY Yang Mills theory also
indicates that the very same perturbation that leads to confinement also
leads to magnetic monopole condensation \cite{seibergwitten}.

There are however many unanswered questions pertaining even to the mere
concept of a magnetic monopole in Yang Mills theories, let alone to the
mechanism of the monopole condensation. Suffice it to say that no reasonable
gauge invariant definition of a monopole is known in pure Yang Mills
theories.

It therefore seems reasonable to keep an open mind on this issue and try and
explore other possibilities. One such alternative mechanism of confinement
was proposed long time ago by t'Hooft \cite{thooft2}. According to it
confinement arises due to condensation of magnetic $Z_{N}$ vortices. These
vortices are interpolated by the t'Hooft disorder operator $V$ which has a
nontrivial commutation relations with the Wilson loop 
\begin{equation}
VW(C)=e^{i{\frac{2\pi }{N_{c}}}n}W(C)V,
\end{equation}
where $n$ is the linking number 
between the (closed) loop $C$ and the curve on
which the disorder operator $V$ is defined (which is a line in $3+1$
dimensions and a point in $2+1$ dimensions). It follows directly from this
commutation relation that if the vacuum of a nonabelian theory is dominated
by configurations with strongly fluctuating number of magnetic vortices
which are only correlated on some finite distance scale, a large Wilson loop
must decay according to the area law.

An apparent problem with this mechanism is that it fails to account for the
properties of the adjoint Wilson loop, and in general of any Wilson loop
defined in a representation with zero $N$-ality. Since the operator $V$
commutes with such a Wilson loop, it does not provide for any explanation of
approximate area law behavior of adjoint Wilson loops that has been observed
in lattice simulations and which is required on general grounds in the large 
$N$ limit\cite{latticeadj}, \cite{greensite1}. Recently however there has
been renewed interest in this scenario due to the results of lattice Monte
Carlo studies which found impressive correlations between confining
properties and the properties of the $Z_{N}$ vortices \cite{tomboulis},\cite
{greensite2}. Some considerations have also been put forward to the effect
that the approximate area law behavior of the adjoint Wilson loop might be
accounted for by finite thickness of the $Z_{N}$ vortices \cite{greensite3}.

Some years ago we have discussed in detail the effective dual Lagrangian
description of $2+1$ nonabelian gauge theories \cite{kovner1}. In this
approach instead of studying a gauge theory in terms of vector potentials,
one considers the Lagrangian for low energy description directly in terms
of the t'Hooft field $V$. This description has some very attractive
features. It is a simple local Lagrangian for a single complex scalar field $%
V$. The detailed structure of the couplings is explicitly calculable in the
weak coupling (partially broken) regime. The charged states in this picture
are topological solitons of the field $V$. The confinement of these states
is understood very simply and vividly on the classical level. It is a direct
and unequivocal consequence of the fact that the dual Lagrangian is not
invariant under the phase rotation of the field $V$, so that even though $V$
has a nonvanishing VEV, the vacuum manifold is not continuous but consists
of discrete number of points. In these circumstances the flux emanating from
a charged state has to flow inside a quasi one - dimensional spatial region
hence producing a flux tube \cite{kovner1}.

Since in $2+1$ the weakly coupled (partially broken Higgs) phase is
connected analytically to the strongly coupled (confinement) phase \cite
{fradkin} one expects that the main features of the dual description are the
same also in the confining region. Quantitatively however the picture in the
confining phase may be somewhat different. It is precisely the purpose of
this paper to extend the dual description to the confining region and to
explore in this language the main qualitative differences between the weakly
coupled and the strongly coupled regime. In this paper we limit ourselves
for simplicity to the $SU(2)$ gauge theory with the Higgs field in adjoint
representation, the so called Georgi - Glashow model. We believe however that
with minimal changes the discussion should also apply to $SU(N)$ theory at any $N$.

The structure of this paper is the following. In Section 2 we recap the
construction of the dual Lagrangian and explain the significance of
different terms in it. We discuss the relation of the parameters in this
Lagrangian with the original couplings of the gauge theory and argue that
certain parameter range corresponds to the limit in which the mass of the
Higgs field becomes infinite. In this strongly coupled regime the dual
Lagrangian should be viewed simply as the effective Lagrangian describing
the dynamics of two lightest glueballs.

In Section 3 we investigate the structure of the fundamental string. We
explicitly construct in the dual theory the Wilson loop operator and show
how to calculate its average. We show that the string tension is equal to
the tension of the domain wall which separates two degenerate vacua which
exist in this theory. This is precisely the picture advocated by t'Hooft a
long time ago \cite{thooft2}. The adjoint string is also considered and
it is shown that, up to some distance scale, the adjoint string indeed has a
nonvanishing string tension. However if the string is too long, it breaks
due to creation of a soliton - antisoliton pair, which in this theory has
finite core energy. It should be noted that in the extreme weak coupling
limit which corresponds to the compact $U(1)$ theory, the core energy of the
soliton becomes infinite (of the order of the ultraviolet cutoff) and the
adjoint string therefore is stable.

In section 4 we study the structure and the interaction between the
fundamental strings. We find that in the weakly coupled regime the
fundamental strings repel each other. The vacuum is therefore reminiscent of
the superconductor of the second kind. The tension of the adjoint string is
twice that of the fundamental one. In general a representation built out of $%
n$ fundamental representation has a string tension approximately 
\begin{equation}
\sigma _{n}=n\sigma _{F}
\end{equation}
where $\sigma _{F}$ is the string tension of the fundamental representation.
This is the same result as found in \cite{greensite4}. Closer to the
confining regime the repulsion between the strings decreases. In fact we
argue that the crossover to the confining regime happens when the
interaction between the strings changes sign. As a consequence the string
tension in this region should behave as 
\begin{equation}
\sigma _{n}=f(n)\sigma _{F},\ \ \ \ \ \ 1<f(n)<n  \label{f1}
\end{equation}
In the extreme situation when the attraction between the strings is large,
the string tension should be independent of the representation, $%
f(n)\rightarrow 1$. It is however unlikely that the pure Yang Mills theory
lies close to this point, and we would expect for pure Yang Mills theory
dependence of the general type eq.(\ref{f1}).

Finally section 5 is devoted to some further discussion.

\section{The effective dual Lagrangian and its relation to the low
energy excitations}

We consider the $SU(2)$ gauge theory with one adjoint Higgs field - the so
called Georgi - Glashow model 
\begin{equation}
{\cal L}=-\frac{1}{4}F_{\mu \nu }^{a}F^{a\mu \nu }+\frac{1}{2}({\cal D}_{\mu
}^{ab}H^{b})^{2}+\tilde{\mu}^{2}H^{2}-\tilde{\lambda}(H^{2})^{2}  \label{lgg}
\end{equation}
where 
\begin{equation}
{\cal D}_{\mu }^{ab}H^{b}=\partial _{\mu }H^{a}-ef^{abc}A_{\mu }^{b}H^{c}
\end{equation}

At large and positive $\tilde{\mu}^{2}$ the model is weakly coupled and
perturbative description is valid. The $SU(2)$ gauge symmetry is broken down
to $U(1)$ and the Higgs mechanism takes place. Two gauge bosons, $W^{\pm }$,
acquire a mass, while the third one, the ``photon'', remains massless
to all orders in perturbation theory. It is well known \cite{polyakov} that
beyond the perturbation theory the photon also acquires an exponentially
small mass $m_{ph}^{2}\propto M_{W}^{2}\exp \{-M_{W}/e^{2}\}$. The same
mechanism leads to confinement of the charged gauge bosons with a tiny
nonperturbative string tension $\sigma \propto e^{2}m_{ph}$.

For negative $\tilde\mu^2$ the gauge symmetry is unbroken, the theory is
strongly interacting and the standard confinement phenomenon is therefore
expected. At large negative $\tilde\mu^2$ the Higgs field becomes heavy and
decouples, and the theory reduces to pure Yang Mills theory. Although the
spectrum in the strong coupling regime and in the weak coupling regime are
different, there is no phase transition between the two regimes but rather a
smooth crossover corresponding to the fact that there is no gauge invariant
order parameter which could distinguish between these two phases \cite
{fradkin}.

We will now recap how the effective dual Lagrangian is constructed in the
weakly coupled phase. First, it is convenient to define a gauge invariant
electric current 
\begin{equation}
J^\mu=\epsilon^{\mu\nu\lambda}\partial_\nu \tilde f_\mu, \ \ \ \ Q=\int d^2
x J_0(x)  \label{qqcd}
\end{equation}
Here 
\begin{equation}
\tilde f_\mu=\epsilon_{\mu\nu\lambda}F^a_{\nu\lambda}\hat H^a
\end{equation}
where $\hat H^a\equiv{\frac{H^a}{|H|}}$. Consider now the following operator 
\begin{equation}
V(x)=\exp {\frac{i}{e}}\int d^2y\ \left[\epsilon_{ij}{\frac{(x-y)_j}{(x-y)^2}%
} \hat{H}^a(y)E^a_i(y)+\Theta(x-y)J_0(y)\right]  \label{VQCD}
\end{equation}
It is the operator of a singular $SU(2)$ gauge transformation with the field
dependent gauge function 
\begin{equation}
\lambda^a(y)={\frac{1}{e}}\Theta(x-y)\hat H^a(\vec y)  \label{lambda}
\end{equation}
This field dependence of the gauge function ensures the gauge
invariance of the operator $V$. 
This is the explicit gauge invariant form of t'Hooft's
``disorder parameter'' \cite{thooft2}. As discussed in detail in \cite
{kovner1}, \cite{kovner2} the operator $V$ is a local scalar field.

It is shown in \cite{kovner1} that the low energy physics of the weakly
coupled phase is conveniently described in terms of the effective Lagrangian
of the field $V$. The general structure of this Lagrangian is determined by
the relevant symmetries very much in the same way as 
the structure of chiral Lagrangian
in QCD is determined by the (spontaneously or explicitly broken) chiral
symmetry. Classically the following current is conserved in this model 
\begin{equation}
\tilde{F}^{\mu }=\tilde{f}^{\mu }-\frac{1}{e}\epsilon ^{\mu \nu \lambda
}\epsilon ^{abc}\hat{H}_{a}({\cal D}_{\nu }\hat{H})^{b}({\cal D}_{\lambda }%
\hat{H})^{c}  \label{F}
\end{equation}
The vortex operator $V$ is a local eigenoperator of
the abelian magnetic field $B(x)=\tilde{F}_{0}$. 
\begin{equation}
\lbrack V(x),B(y)]=-{\frac{2\pi }{e}}V(x)\delta ^{2}(x-y)  \label{com}
\end{equation}
That is to say, when acting on a state it creates a pointlike magnetic vortex
which carries a quantized unit of magnetic flux. The transformation
generated by the magnetic flux $\Phi \equiv \int d^{2}xB(x)$ therefore acts
on the vortex field $V$ as a phase rotation 
\begin{equation}
e^{i\alpha \Phi }V(x)e^{-i\alpha \Phi }=e^{i{\frac{2\pi \alpha }{e}}}V(x)
\label{magntr}
\end{equation}

The classical conservation of $\Phi$ is spoiled quantum mechanically by the
presence of t'Hooft-Polyakov monopoles. However the discreet $Z_2$ subgroup
of the transformation group eq.(\ref{magntr}) $V\rightarrow -V$ remains
unbroken in the quantum theory as well \cite{kovner1}. The effective
Lagrangian preserves this $Z_2$ symmetry. Its form is 
\begin{equation}
{\cal L}=\partial_\mu V^*\partial^\mu V -\lambda(V^*V-\mu^2)^2 -{\frac{m^2}{4%
}}(V^2 +V^{*2}) + \zeta(\epsilon_{\mu\nu\lambda}\partial_\nu V^*
\partial_\lambda V)^2  \label{ldualgg}
\end{equation}
The coupling constants in eq.(\ref{ldualgg}) are determined in the weakly
coupled region from perturbation theory and dilute monopole gas
approximation. In the weakly coupled region we have 
\begin{eqnarray}  \label{couplings}
&&\mu^2={\frac{e^2}{8\pi^2}}  \nonumber \\
&&\lambda={\frac{2\pi^2 M^2_H}{e^2}} \\
&&m=m_{ph}  \nonumber \\
&&\zeta\propto {\frac{M_W}{e^4M_H^2}}  \nonumber
\end{eqnarray}
Here $M_H$ is the Higgs mass in the original theory, $M_W$ is the mass of
the $W$-boson and $m_{ph}$ is the exponentially small nonperturbative photon
mass calculated by Polyakov \cite{polyakov}.

The first term in this Lagrangian is the standard kinetic term for the field 
$V$. The second term is a potential invariant under the $U(1)$ phase
rotation. As discussed above, this rotation is generated by the magnetic
flux. The third term is the manifestation of the quantum mechanical breaking
of the flux symmetry and its coefficient is proportional to the monopole
density, hence large suppression in the weak coupling limit. Finally the
last term contains four derivatives. It is the analog of the Skyrme
term
in the effective chiral Lagrangian of QCD. As
befits a higher derivative term it is not relevant if we wish to discuss
physics only at scales below the mass of $W^{\pm }$. We chose to keep it
since we want discuss the issue of the string breaking which as we shall
see is sensitive to this term.

Let us discuss some simple properties of the dual description. First, we are
only going to consider the model for $\mu ^{2}>0$, and therefore vacuum
expectation value of $V$ is nonvanishing. In fact according to eq.(\ref
{couplings}) this VEV is proportional to the gauge coupling constant. To
understand why this is so let us recall how a charged state is
represented
in this
dual description. As shown in \cite{kovner1} the electric current
eq.(\ref{qqcd}) can be expressed in terms of the vortex operator eq.(\ref
{VQCD}) in the form 
\begin{equation}
{\frac{e}{\pi }}J_{\mu }=i\epsilon _{\mu \nu \lambda }\partial _{\nu
}(V^{\ast }\partial _{\lambda }V)
\end{equation}
The electric charge is therefore proportional to the winding number of
the phase of the field $V$. 
A charge state is a soliton of $V$ with a nonzero winding number.
Neglecting for a moment the small 
$U(1)$ noninvariant term in eq.(\ref{ldualgg}), we find that the minimal
energy configuration in one soliton sector is a rotationally invariant
hedgehog, Fig.1, which far from the soliton core has the form 
\begin{equation}
V(x)=\mu e^{i\theta (x)}.
\end{equation}
Here $\theta (x)$ is an angle between the vector $x$  and one of the
axes. 
\begin{figure}
\begin{center}
\leavevmode
\epsfxsize=5.3in
\epsfbox{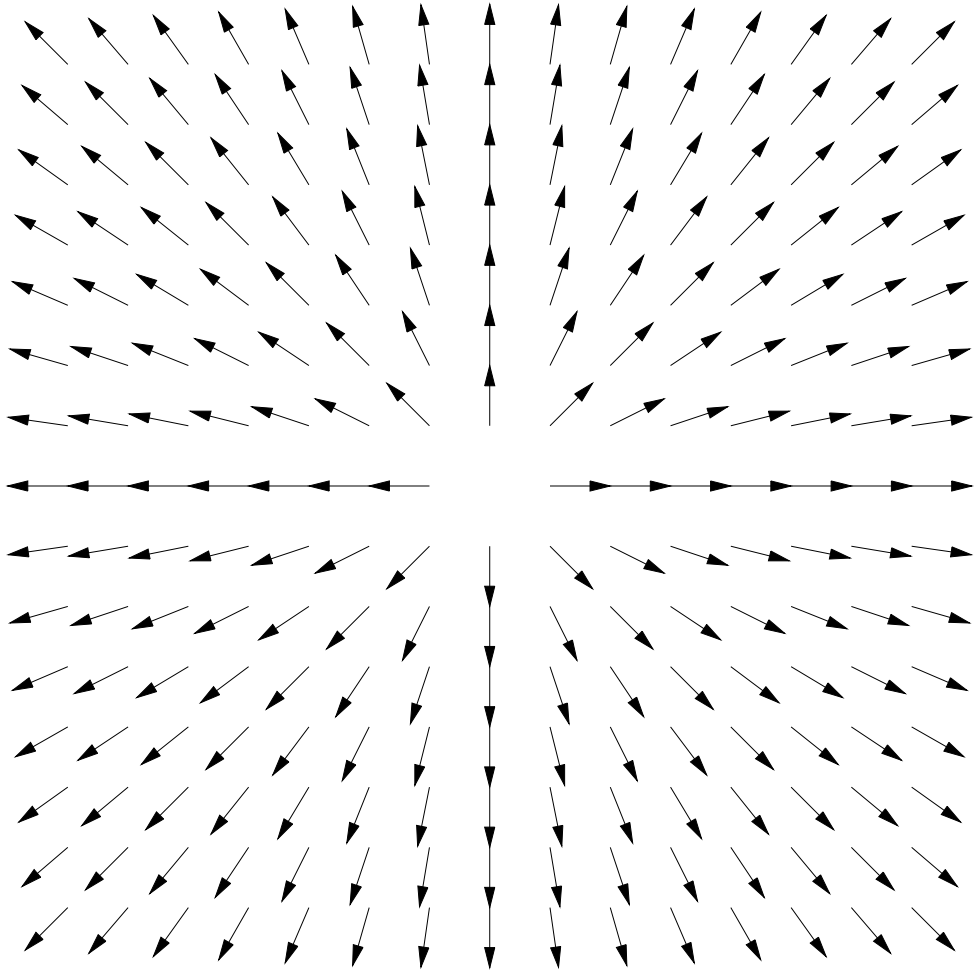}
\caption{The hedgehog configuration of the field $V$
in the state of unit charge when the symmetry breaking terms in the
effective Lagrangian are neglected.}
\label{fig1a}
\end{center}
\end{figure}
The self energy of this configuration is logarithmically divergent in
the infrared due to the contribution from the kinetic term 
\begin{equation}
E=2\pi \mu ^{2}\ln M_{H}L
\end{equation}
This self energy should be equal to the electromagnetic energy of a charged
state in the original description. The Coulomb part of the self energy of a
charged state indeed diverges logarithmically. Matching the coefficients of
the logarithmic divergence gives the first of the equations in eq.(\ref
{couplings}).

In this argument we have neglected the $U(1)$ noninvariant term in the
Lagrangian, since its coefficient is nonperturbatively small. Without this
term the Lagrangian eq.(\ref{ldualgg}) contains a massless excitation - the
phase of the field $V$. This is of course the photon which is indeed
massless in perturbation theory \footnote{%
Note that in 2+1 dimensions a massless photon is a scalar particle.}. It is
massless since the theory has infinite number of vacua, corresponding
to an
arbitrary phase of the VEV of $V$. Due to this the energy of a charged
soliton is only logarithmically divergent and there is no linear
confinement.

If we now reinstate the symmetry breaking term the situation changes
qualitatively. First of all the degeneracy between the different phases of $V
$ is lifted and the vacuum manifold now consists of only two points $V=\pm
\mu $. Consequently, the theory is not a gapless one anymore. The lightest
excitation of the 
Lagrangian eq.(\ref{ldualgg}) is still the phase of $V$ which
now has a small mass $m$. This field also becomes self interacting with
the sine-Gordon type potential which follows directly from the $U(1)$
noninvariant term in eq.(\ref{ldualgg}). This is precisely the self -
interaction of the massive photon calculated first in \cite{polyakov}.

The explicit symmetry breaking causes a more dramatic change in the
topologically charged (soliton) sector. The energy of an isolated soliton
now diverges in the infrared {\it linearly} rather than just
logarithmically. The configuration that forms around the soliton is a
string, Fig.2, rather than a hedgehog of Fig.1. The hedgehog
configuration is no longer energetically favoured because in such a
configuration the phase of the field $V$ is far from its vacuum value almost
everywhere in space. The energy of a hedgehog therefore becomes
quadratically divergent: $E\propto e^{2}m^{2}L^{2}$, where $L$ is an
infrared cutoff. To minimize the energy for a nonzero winding, the system
chooses a stringlike configuration Fig.2. The phase of $V(x)$ deviates
from $0$ (or $\pi $) only inside a strip of width $d\sim 1/m$ stretching
from the location of the defect to infinity. The energy of such a
configuration is linearly divergent. This is the simple picture of
confinement in the dual formulation in the weakly coupled regime.

\begin{figure}
\begin{center}
\leavevmode
\epsfxsize=5.3in
\epsfbox{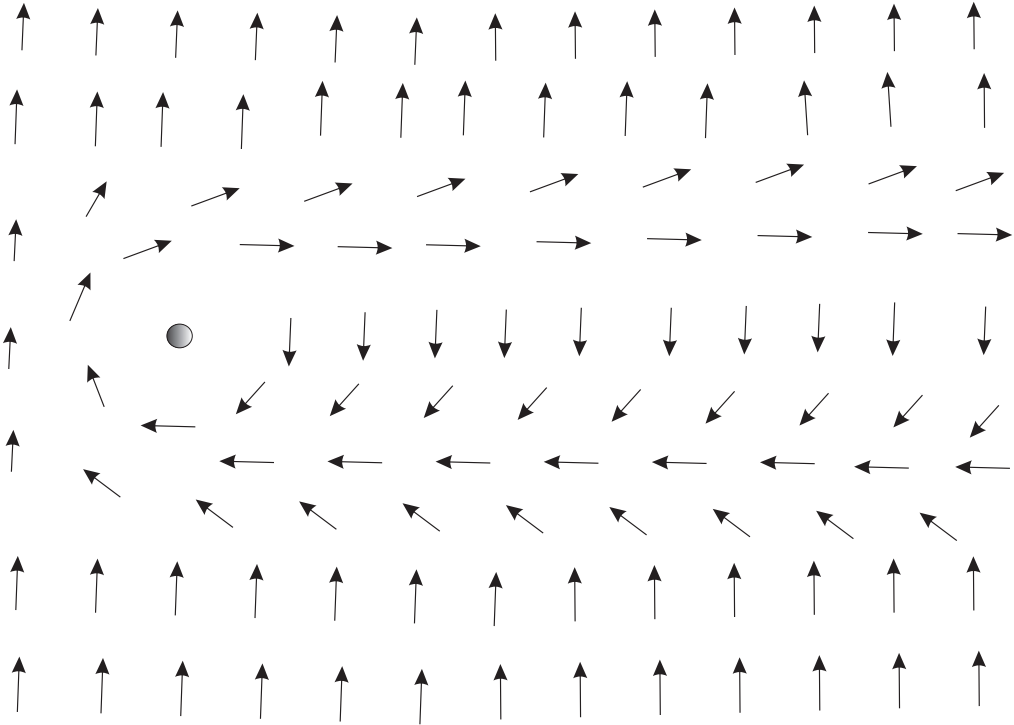}
\caption{The  stringlike configuration of the field $V$
in the state of unit charge in the presence of 
the symmetry breaking terms in the
effective Lagrangian.}
\label{fig1b}
\end{center}
\end{figure}

We shall see in the next section that this discussion is appropriate for the
adjoint string. The fundamental string appears in a somewhat different
fashion. 

Before discussing  the confining properties of
the theory in more detail let us
make the following comment concerning the relation between fields entering
the effective Lagrangian and physical excitations of the nonabelian theory. 

The vortex operator as defined in eq.(\ref{VQCD}) has a fixed length whereas
the field $V$ which enters the Lagrangian eq.(\ref{ldualgg}) is a
conventional complex field. How should one understand that? First of
all at weak gauge coupling the quartic coupling in the dual Lagrangian
is large $\lambda \rightarrow \infty $. This condition
freezes the radius of $V$ dynamically. In fact even at finite value of $%
\lambda $ if one is interested in the low energy physics, the radial
component is irrelevant as long as it is much heavier than the phase. Indeed
at weak gauge 
coupling the phase of $V$ which interpolates the massive photon is
much lighter than all the other excitations in the theory. Effectively
therefore at low energies eq.(\ref{ldualgg}) reduces to a nonlinear sigma
model and one can identify the field $V$ entering eq.(\ref{ldualgg})
directly with the vortex operator of eq.(\ref{VQCD}). However it is well
known that quantum mechanically the radial degree of freedom of a sigma
model field is always resurrected. The spectrum of such a theory always
contains a scalar particle which can be combined with the phase into a
variable length complex field. The question is only quantitative - how heavy
is this scalar field relative to the phase.

Another way of expressing this is the following. The fixed length field $V$
is defined at the scale of the UV cutoff in the original theory. To arrive
at the low energy effective Lagrangian one has to integrate over all quantum
fluctuations down to some much lower energy scale. In the process of this
integrating out the field is ``renormalized'' and it acquires a dynamical
radial part. The mass of this radial part then is just equal to the mass of
the lowest particle with the same quantum numbers in the original theory.
This is in fact why the parameters in eq.(\ref{couplings}) are such that the
mass of the radial part of $V$ is equal to the mass of the scalar Higgs
particle.

This brings us to the following observation. We know how to calculate the
couplings of the dual Lagrangian only in the weak gauge 
coupling limit. However we
also know that the weak and the strong coupling regimes in this model are
not separated by a phase transition. It is therefore plausible that the low
energy dynamics at strong coupling is described by the same effective
Lagrangian. Clearly if this is the case the degrees of freedom that enter
this Lagrangian must interpolate real low energy physical states of the
strong coupling regime, that is to say must correspond to lightest glueballs
of pure $SU(2)$ Yang Mills theory. It turns out that in fact the quantum
numbers of two lightest glueballs are precisely the same as those of the
phase and the radius of the vortex field $V$. The radial part of $V$ is
obviously a scalar and has quantum numbers $0^{++}$. The quantum numbers of
the phase are easily determined from the definition eq.(\ref{VQCD}). Those
are $0^{--}$. The spectrum of pure $SU(N)$ Yang Mills theory in 2+1
dimensions was extensively studied recently on the lattice \cite{teper}. The
two lightest glueballs for any $N$ are found to have exactly those quantum
numbers. The lightest excitation is the scalar while the next one is a
charge conjugation odd pseudoscalar with the ratio of the masses roughly $%
m_{p}/m_{s}=1.5$ for any $N$\footnote{%
Actually this state of matters is firmly established only for $N>2$. At $N=2$
the mass of the pseudoscalar has not been calculated in \cite{teper}. The
reason is that it is not clear how to construct a charge conjugation odd
operator in a pure gauge $SU(2)$ lattice theory. So it is possible that the
situation at $N=2$ is nongeneric in this respect. In this case we would hope
that our strong coupling picture applies at $N>2$.}.

We are therefore lead to the following conjecture. The low energy physics of
the $SU(2)$ gauge theory is always described by the effective Lagrangian eq.(%
\ref{ldualgg}). In the weak coupling regime the parameters are given in eq.(%
\ref{couplings}). Here the pseudoscalar particle is the lightest and the
scalar is the first excitation. The pseudoscalar
as an almost massless photon and the scalar is the massive Higgs
particle. Moving towards the strong coupling regime (decreasing the Higgs
VEV in the original language) corresponds to increasing the pseudoscalar
mass while reducing the scalar mass and the parameters of the effective
Lagrangian change accordingly. The crossover between the weak and the strong
coupling regimes occurs roughly where the scalar and the pseudoscalar become
degenerate. At strong coupling the degrees
of freedom in the effective Lagrangian are the two lightest glueballs. They
are however still collected in one complex field which represents
nontrivially the exact $Z_{2}$ symmetry of the theory. We stress
that the existence of the $Z_{2}$ symmetry in this model is an exact
statement which is not related to the weak coupling limit. It is therefore
completely natural to expect that this symmetry must be nontrivially
represented in the effective low energy Lagrangian. Of course the spectrum
of pure Yang Mills theory apart from a scalar and a pseudoscalar glueballs
contains many other massive glueball states and those are not separated by a
large gap from the two lowest ones. Application of this effective Lagrangian
to the strong coupling regime therefore has to be taken in a qualitative
sense.

In the rest of this paper we will explore the consequences of this picture
on the properties of confinement. We will see that although in both
regimes the
confining strings arise naturally in the effective Lagrangian
description, 
their
properties depend in an important way on whether the lightest particle is a
scalar or a pseudoscalar.

\section{String tension and the breaking of the adjoint string.}

We now turn to the study of the
confining properties of the theory. In the preceding
section we have briefly described why the charged states have linearly IR
divergent energy. In this section the discussion will be made slightly more
formal. First we want to discuss the fundamental string and its string
tension. For this we need to know how to calculate the expectation value of
a fundamental Wilson loop in the dual theory. We therefore start
by constructing the Wilson loop operator.

Let us first consider a spacelike Wilson loop. The vortex operator eq.(\ref
{VQCD}) and the fundamental Wilson loop operator satisfy the t'Hooft
commutation relation 
\begin{equation}
W(C)V(x)=V(x)W(C)e^{i\pi n(C,x)}  \label{comm}
\end{equation}
where $n(C,x)$ is the linking number between the loop $C$ and the point $x$: 
$n(C,x)=1$ if $x$ is inside the surface bounded by $C$ and vanishes
otherwise. To see this use the fact that under a gauge transformation
the Wilson loop transforms as 
\begin{equation}
U^{\dagger }W(C)U=W(C)Pe^{\int_{C}dl_{i}U^{\dagger }\partial _{i}U}
\label{gauge}
\end{equation}
For regular gauge functions $U(x)$ the phase factor on the RHS vanishes and
the Wilson loop is invariant. The operator $V$ however is the operator
of a singular gauge transformation. The gauge function $\lambda ^{a}$ eq.(%
\ref{lambda}) has a discontinuity equal to $\pi $ along the cut of the
angular function $\theta (x)$. For this function the phase factor in eq.(\ref
{gauge}) is unity if the cut crosses the contour $C$ an even number of times
and equals $-1$ when the cut crosses $C$ an odd number of times. This leads
to the commutator eq.(\ref{comm}).

Now it is straightforward to write down an operator in terms of the field $V$
that has the same property 
\begin{equation}
W(C)=e^{i\pi \int_{S}d^{2}xP(x)}  \label{wld}
\end{equation}
Here $S$ is the surface bounded by the contour $C$, and $P$ is the operator
of momentum conjugate to the phase of $V$. Introducing the radius and the
phase of $V$ by 
\begin{equation}
V(x)=\rho (x)e^{i\chi (x)}
\end{equation}
one can write the path integral representation for calculating the vacuum
average of the Wilson loop as 
\begin{equation}
<W(C)>=\int DV\exp \left\{i\int d^{3}x\rho ^{2}(\partial _{\mu }\chi -j_{\mu
}^{S})^{2}+(\partial _{\mu }\rho )^{2}-U(V)\right\}  
\label{wldpi}
\end{equation}
where $U(V)$ is the $Z_{2}$ invariant potential of eq.(\ref{ldualgg}). The
external current $j_{\mu }^{S}(x)$ does not vanish only at points $x$ which
belong to the surface $S$ and is proportional to the unit normal $n_{\mu }$
to the surface S. Its magnitude is such that when integrated in the
direction of $n$ it is equal to $\pi $. These properties are
conveniently
encoded in the following expression
\begin{equation}
\int_{T}dx_{\mu }j_{\mu }^{S}(x)=\pi n(T,C)  \label{angle}
\end{equation}
Here $T$ is an arbitrary closed contour, and $n(T,C)$ is the linking number
between two closed curves $T$ and $C$. In eq.(\ref{wldpi})
we have neglected the four
derivative term present in 
eq.(\ref{ldualgg}) for simplicity. Have we kept it, the
derivative of the phase field $\partial _{\mu }\chi $ would have been
shifted by the same current $j^S_{\mu }$ in this term also. The expression eq.(%
\ref{wldpi}) follows directly from the operatorial definition of the Wilson
loop eq.(\ref{wld}), if one properly takes care of the seagull terms. Those
are responsible for the appearance of the $j_{\mu }^{2}$ term in eq.(\ref
{wldpi}). This issue for operators similar to (\ref{wld}) is discussed in
detail in \cite{kovner2} and \cite{frohlich}.

So far we have considered spatial Wilson loops. However the expression eq.(%
\ref{wldpi}) is completely covariant, and in this form is valid for timelike
Wilson loops as well. It is important to note that although the expression
for the current depends on the surface $S$, the Wilson loop operator in fact
depends only on the contour $C$ that bounds this surface. A simple way to
see this is to observe that a change of variables
$\chi\rightarrow\chi+\pi$ 
in the
volume bounded by $S+S^{\prime}$ leads to the change $j_\mu^S\rightarrow
j_\mu^{S^{\prime}}$ in eq.(\ref{wldpi}). The potential is not affected by
this change since it is globally $Z_2$ invariant. Therefore the operators
defined with $S$ and $S^{\prime}$ are completely equivalent.

To calculate the energy of a pair of static fundamental charges at
points $A$ and $B$ we have to consider a timelike fundamental Wilson loop of
infinite time dimension. This corresponds to time independent $j_\mu$ which
does not vanish only along a spatial curve $G$ connecting the two points and
pointing in the direction normal to this curve Fig.3. 
The shape of the curve
itself does not matter, since changing the curve without changing its
endpoints is equivalent to changing the surface $S$ in eq.(\ref{wldpi}).
\begin{figure}
\begin{center}
\leavevmode
\epsfxsize=3.3in
\epsfbox{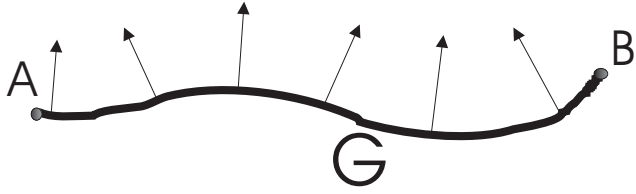}
\caption{The external current $j_\mu$
which creates the pair of static fundamental charges 
in the effective Lagrangian description.}
\label{fig2}
\end{center}
\end{figure}
In the classical approximation the path integral eq.(\ref{wldpi}) is
dominated by a static configuration of $V$. To determine it we have to
minimize the energy on static configurations in the presence of the external
current $j_{\mu }$. The qualitative features of the minimal energy solution
are quite clear. The effect of the external current clearly is to flip the
phase of $V$ by $\pi $ across the curve $G$, as is expressed in eq. (\ref
{angle}). Any configuration that does not have this behavior will have the
energy proportional to the length of $G$ and to the UV cutoff scale.
Recall that the vacuum in our theory is doubly degenerate. The
sign change 
of $V$ transforms one vacuum configuration into the other one. The
presence of $j_{\mu }$ therefore requires that on opposite sides of the
curve $G$, immediately adjacent to $G$ there should be different vacuum
states. It is clear however that far away from $G$ in either direction the
field should approach the same vacuum state, otherwise the energy of a
configuration diverges linearly in the infrared. The phase of $V$ therefore
has to make half a wind somewhere in space to return to the same vacuum
state far below $G$ as the one that exists far above $G$. If the distance
between $A$ and $B$ is much larger than the mass of the lightest particle in
the theory, this is achieved by having a segment of a domain wall between
the two vacua connecting the points $A$ and $B$. Clearly to minimize the
energy the domain wall must connect $A$ and $B$ along a straight line.
The energy of such a domain wall is proportional to its length, and
therefore the Wilson loop has an area law behavior. The minimal energy
solution is schematically depicted on Fig.4.
\begin{figure}
\begin{center}
\leavevmode
\epsfxsize=5.3in
\epsfbox{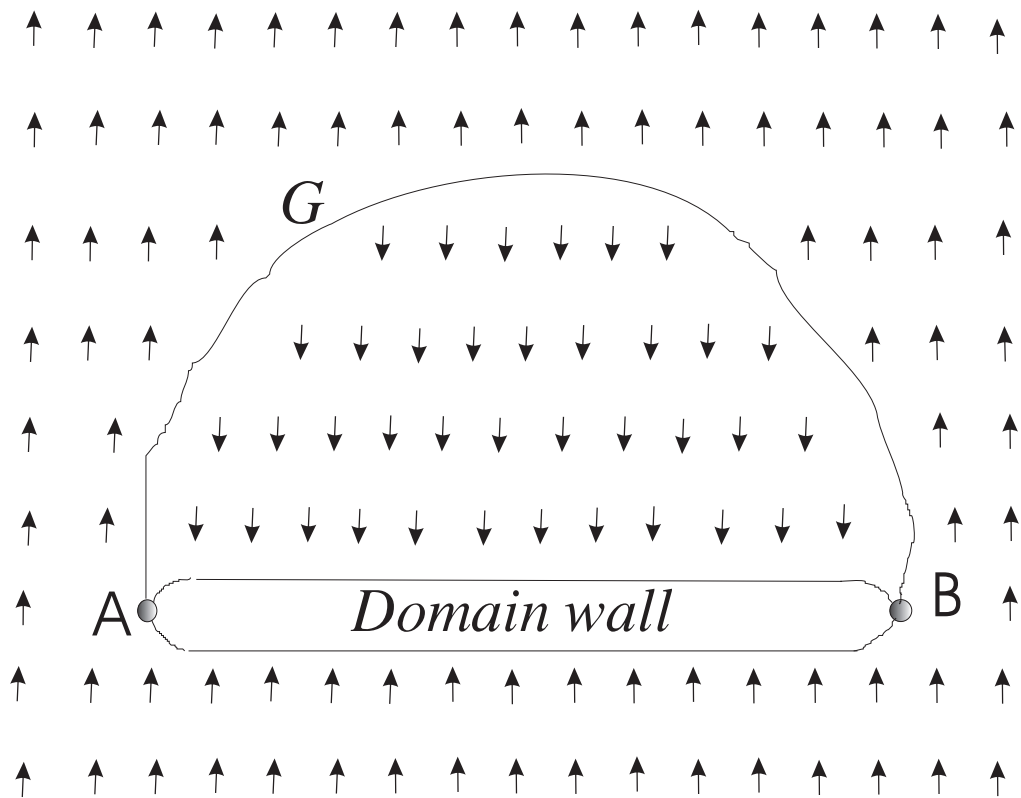}
\caption{The minimal energy configuration of $V$ in the presence of a
pair of fundamental charges.}
\label{fig3}
\end{center}
\end{figure}
We see that the string tension for the fundamental string is equal to the
tension of the domain wall which separates the two vacua in the theory. This
conclusion is equally valid in the weakly and strongly coupled regimes. The
structure of the domain wall and the relation between the tension and the
parameters of the theory is quite different in the two regimes and we will
come back to this question in the next section.

Note that the fundamental string is an absolutely stable topological
object in the $Z_{2}$ invariant theory: the domain wall. It can not break,
if one makes the distance between the two charges larger. From the point of
view of the original theory this is so because the theory does not contain
particles with fundamental charge. In the dual description it is also
obvious since there is no object in the theory on which a domain wall can
terminate. One way of thinking about it is in terms of the ''electric
charge'' eq.(\ref{qqcd}). As discussed in the previous section this charge
counts the number of windings of the field $V$. Across the domain wall the
phase of $V$ changes by $\pi $, therefore the winding number of a point at
which the domain wall can end must be half integer. Finite energy states
with half integer winding do not exist in the theory and therefore
the fundamental string can not break. The only reason why the wall can end
on a fundamental external charge is because of the external current $j_{\mu }$ which
itself furnishes an extra half a wind, so that the total winding number
around points $A$ and $B$ is still integer.

The situation is rather different if we consider adjoint external
charges instead. Here we have to be more precise what we mean by that. A
fundamental Wilson loop gives an energy of a state with two heavy external
charges in a fundamental representation. We would now like to introduce two
external charges in adjoint representation. The adjoint Wilson loop however
has a trivial commutation relation with the vortex operator, and this
therefore does not tell us what operator we should consider in the dual
theory in order to ''measure'' the interaction energy of the sources. 
Moreover in the
weakly coupled phase, where the gauge symmetry is ''broken'' not all
components of the external adjoint charge are equivalent. The three states
in the adjoint representation of $SU(2)$ split into three representation
with respect to the unbroken electric charge generator eq.({\ref{qqcd}),
with eigenvalues $0,1$ and $-1$\footnote{%
Note that for the fundamental external charge $Q=\pm 1/2$.}. The
interactions of these three states are obviously different. The neutral
state does not couple directly to the photon. It is therefore not expected
to feel any confining force at all and is not interesting from our point of
view. The remaining two states carry electric charge and interact just like $%
W^{\pm }$ through the photon exchange. We therefore would like to
concentrate on these states. In the following whenever we speak about the
adjoint string we mean the string between the external charges $1$ and $-1$. 
}

The introduction of an external charge $Q=1$ at a point $A$ from the point
of view of the dual theory means that we force the theory into the
topological sector with unit winding of $V$ around $A$. The dipole
configuration corresponds to a unit wind at $A$ and a minus one wind at $B$.

What is the minimum energy configuration in this sector? There are two types
of configurations which should be considered (we are assuming that the
distance between the external charges is much larger than the inverse mass
of the lowest excitation). First there is a stringlike configuration, where
the points $A$ and $B$ are connected by a ''double domain wall'', inside
which the phase of the field $V$ completes a rotation by $2\pi $. Thus here
the flux emanating from $A$ propagates all the way to $B$ and terminates
there. This is depicted schematically on Fig.5. These are two charges of
Fig. 2 joined by a string.
\begin{figure}
\begin{center}
\leavevmode
\epsfxsize=6.3in
\epsfbox{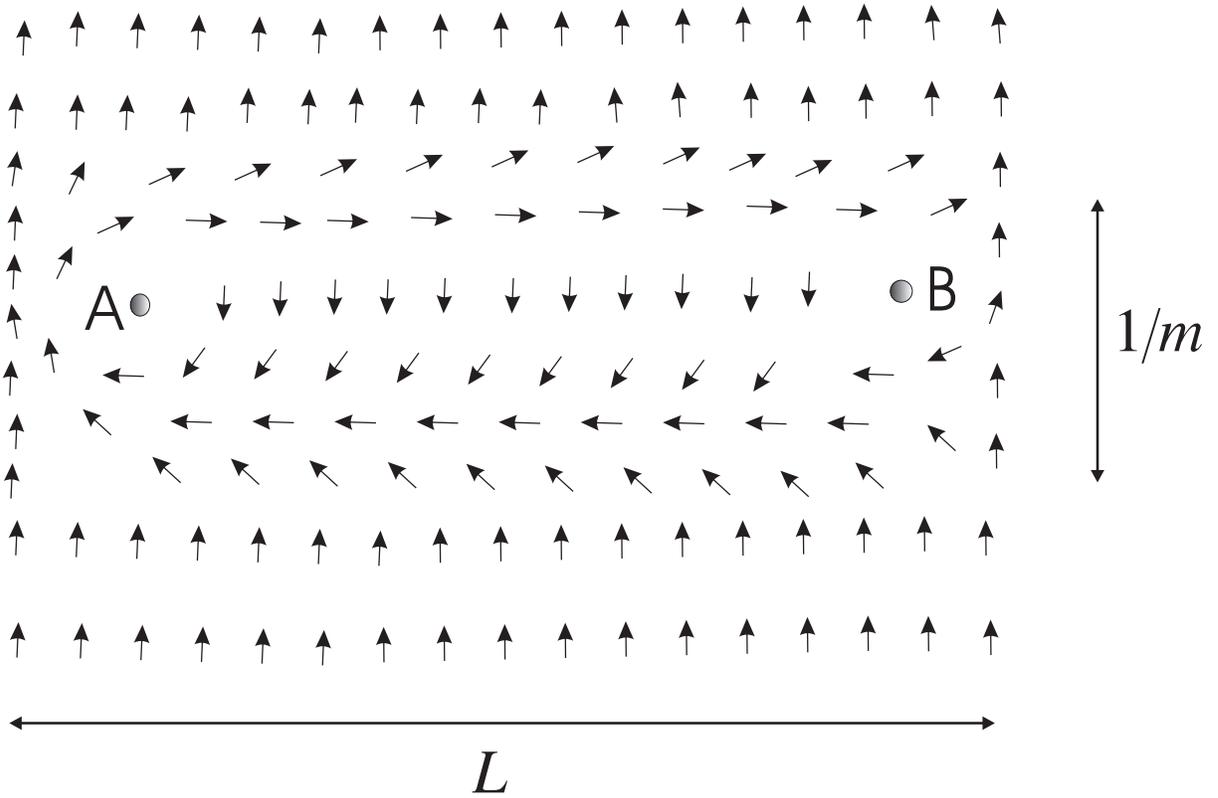}
\caption{The stringlike configuration of $V$ in the presence of a
pair of adjoint charges. This configuration has the minimal energy if $L<L_c$.}
\label{fig4}
\end{center}
\end{figure}
The energy of this configuration depends linearly on the distance $L$. The
energy density per unit length is obviously of the same order as that for
the domain wall, although somewhat higher. The energy of this type of
configuration is therefore 
\begin{equation}
E_1=x\sigma_FL
\end{equation}
Here $x$ is a number of order one, and $\sigma_F$ - the fundamental string
tension.

Another type of configuration is where the flux emanating from $A$ is
screened locally by creating a topological soliton nearby. This is possible
since in order to screen the flux the soliton has to carry an integer
winding number rather than halfinteger as in the case of the fundamental
charge. There are such solitonic solutions of classical equations of motion
of eq.(\ref{ldualgg}). The structure of these solutions is not difficult to
understand. The radial field $\rho $ must vanish in the core of such a
configuration. The size of the core, that is the area in which $\rho $
significantly differs from its value in the vacuum must be of order $%
S=1/M_{H}^{2}$. The charge density within this area is of order $J_{0}\sim
\mu ^{2}S=e^{2}M_{H}^{2}$. The energy associated with the core comes mainly
from the four derivative term in eq.(\ref{ldualgg}) and is 
\begin{equation}
E_{c}\sim \zeta e^{4}M_{H}^{2}
\end{equation}

The energy of the screened dipole configuration is given by (neglecting the
Coulomb energy of the two dipoles) 
\begin{equation}
E_2=2E_c
\end{equation}

From this analysis it is clear that as long as the distance
between the external charges does not exceed the critical value $L_{c}\sim
2\zeta e^{4}M_{H}^{2}/\sigma _{F}$ the energetically favourable option is a
string. However at larger distances the string breaks and it is the screened
dipole configuration that has lower energy.

The solitonic solution of the dual Lagrangian obviously has to be identified
with the lightest charged particle in the Nonabelian gauge theory. In the
weakly coupled regime this is just $W^\pm$. Indeed with the coupling $\zeta$
given by eq.(\ref{couplings}) the core energy of the soliton is equal to the
mass of the $W^\pm$. 

At weak coupling the mass of the charged boson is
related to the VEV of the Higgs field by 
\begin{equation}
M_W\sim e<H>
\end{equation}
On the other hand the string tension is determined by the mass of the photon 
\cite{polyakov} 
\begin{equation}
\sigma_F\sim e^2m
\end{equation}
We therefore have 
\begin{equation}
L_c\sim {\frac{<H>}{e}}{\frac{1}{m}}>>{\frac{1}{m}}
\end{equation}
In this regime therefore the critical length is much larger than the inverse
photon mass, which determines the thickness of the string. Nevertheless if
the string is too long it breaks. It has been emphasized recently in \cite
{greensite4} that the breaking of the string is a qualitative phenomenon
which distinguishes the Georgi - Glashow model from the compact $U(1)$
theory. In the latter all strings are stable. In the dual Lagrangian
approach this is indeed seen very clearly. The compact $U(1)$ theory limit
is obtained from eq.(\ref{ldualgg}) at $\lambda\rightarrow\infty$, $%
\zeta\rightarrow\infty$. In this limit the core energy of the soliton
diverges and the critical length becomes infinite.

In the weak coupling regime 
one can reasonably speak about a well developed string long before it
breaks. In the strong coupling regime the situation is less clear. We will
briefly come back to this point in the next section.

\section{Strong versus weak coupling and the string interaction.}

In this section we discuss the structure of the fundamental string and then
the interaction between the strings. Consider first the weakly coupled
regime. As discussed above in terms of the coupling constants in the
effective Lagrangian this corresponds to the limit where the phase of $V$ is
much lighter than the radial part. A cartoon of the fundamental string in
this situation is depicted in Fig.6.
\begin{figure}
\begin{center}
\leavevmode
\epsfxsize=6.3in
\epsfbox{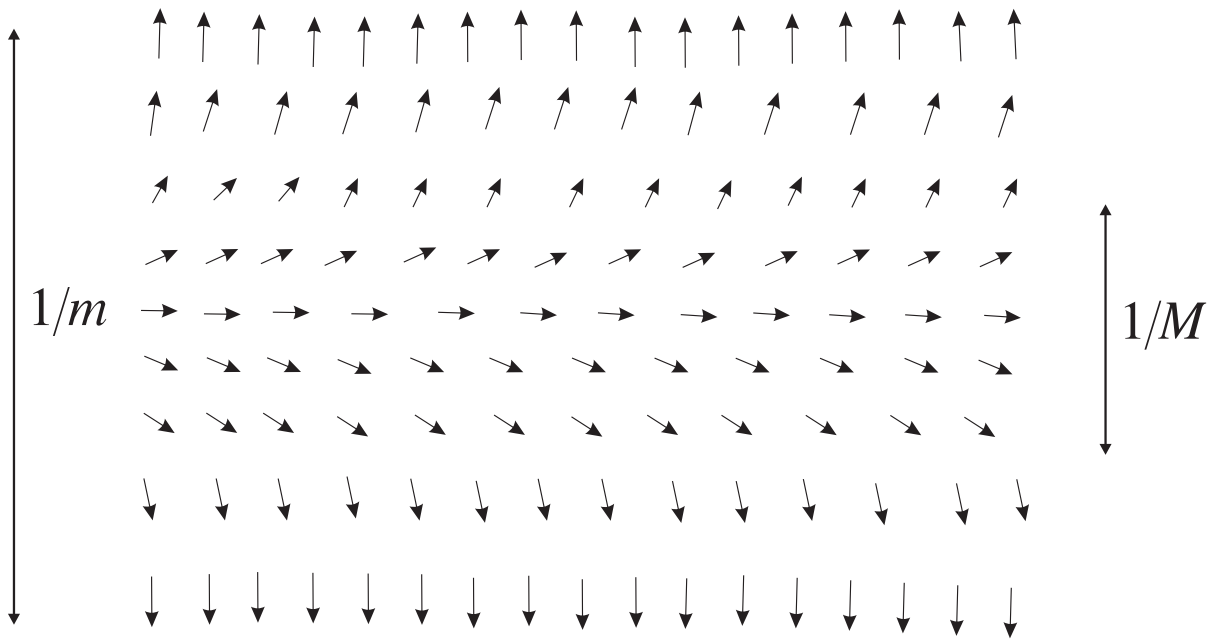}
\caption{The structure of the string (domain wall) in the regime when
the pseudoscalar is lighter than the scalar, $m<M$.}
\label{fig5}
\end{center}
\end{figure}
The radial part $\rho$ being very heavy practically does not change inside
the string. In fact the value of $\rho$ in the middle of the string can
be estimated from the following simple argument. The width of region where $%
\rho$ varies from its vacuum value $\mu$ to the value $\rho_0$ in the middle
is of the order of the inverse mass of $\rho$. The energy per unit length
that this variation costs is 
\begin{equation}
\sigma_\rho\sim M(\mu-\rho_0)^2+{\frac{m^2}{M}}\rho_0^2
\end{equation}
where the first term is the contribution of the kinetic term and the second
contribution comes from the interaction term between $\rho$ and $\chi$ due
to the fact that the value of $\chi$ in the middle of the string differs
from its vacuum value. Minimizing this with respect to $\rho_0$ we find 
\begin{equation}
\rho_0=\mu (1-x{\frac{m^2}{M^2}})
\end{equation}
We see therefore that even in the middle of the string the difference in the
value of $\rho$ and its VEV is second order in the small ratio $m/M$.
Correspondingly the contribution of the energy density of $\rho$ to the
total energy density is also very small. 
\begin{equation}
\sigma_\rho\sim {\frac{m }{M}}m\mu^2
\end{equation}
This is to be compared with the total tension of the string which is
contributed mainly by the pseudoscalar phase $\chi$ 
\begin{equation}
\sigma_\chi\sim m\mu^2
\end{equation}
This again we obtain by estimating the kinetic energy of $\chi$ on a
configuration of width $1/m$ where $\chi$ changes by an amount of order $1
$.

Remembering that $\mu^2\sim e^2$ we see that the fundamental string tension
parametrically is 
\begin{equation}
\sigma_F\sim m e^2
\end{equation}

This is consistent with the Polyakov's calculation in dilute monopole gas
approximation \cite{polyakov}.

It is worth stressing the following important feature of this simple
analysis. The heavy radial field $\rho $ practically does not contribute to
the string tension. This is natural of course from the point of view of
decoupling. In the limit of infinite mass $\rho $ should decouple from the
theory without changing its physical properties. It is however very
different from the situation in superconductors. In a superconductor of the
second kind, where the order parameter field is much heavier than the
photon (correlation length is smaller than the penetration depth $\kappa >
\frac{1}{\sqrt{2}}$) the magnetic field and the order parameter give
contributions of the same order to the energy density of the Abrikosov
vortex (up to logarithmic corrections $O(\log \kappa) $). 
This is the consequence of the fact
that the order parameter itself is forced to vanish in the core of the
Abrikosov vortex, and therefore even though it is heavy, its variation inside
the vortex is large. An even more spectacular situation arises if we
consider a domain wall between two vacuum states in which the heavy field
has different values \cite{kovner3}. In this situation the contribution of
the heavy field $\phi$ to the tension would be 
\begin{equation}
\sigma _{heavy}=M(\Delta \phi)^{2}
\end{equation}
where $\Delta \phi$ is the difference in the values of $\phi$ 
on both sides of the
wall. For fixed $\Delta \phi$ the energy density diverges when $\phi$ 
becomes
heavy. In our case this does not happen since the two vacua which are
separated by the domain wall differ only in VEV of the light field $\chi $
and not the heavy field $\rho $.

Let us now consider the fundamental string in the opposite regime, that is
when the mass of the scalar is much smaller than the mass of the
pseudoscalar. The profile of the fields in the wall now is very different.
The cartoon of this situation is given on Fig.7.
\begin{figure}
\begin{center}
\leavevmode
\epsfxsize=6.3in
\epsfbox{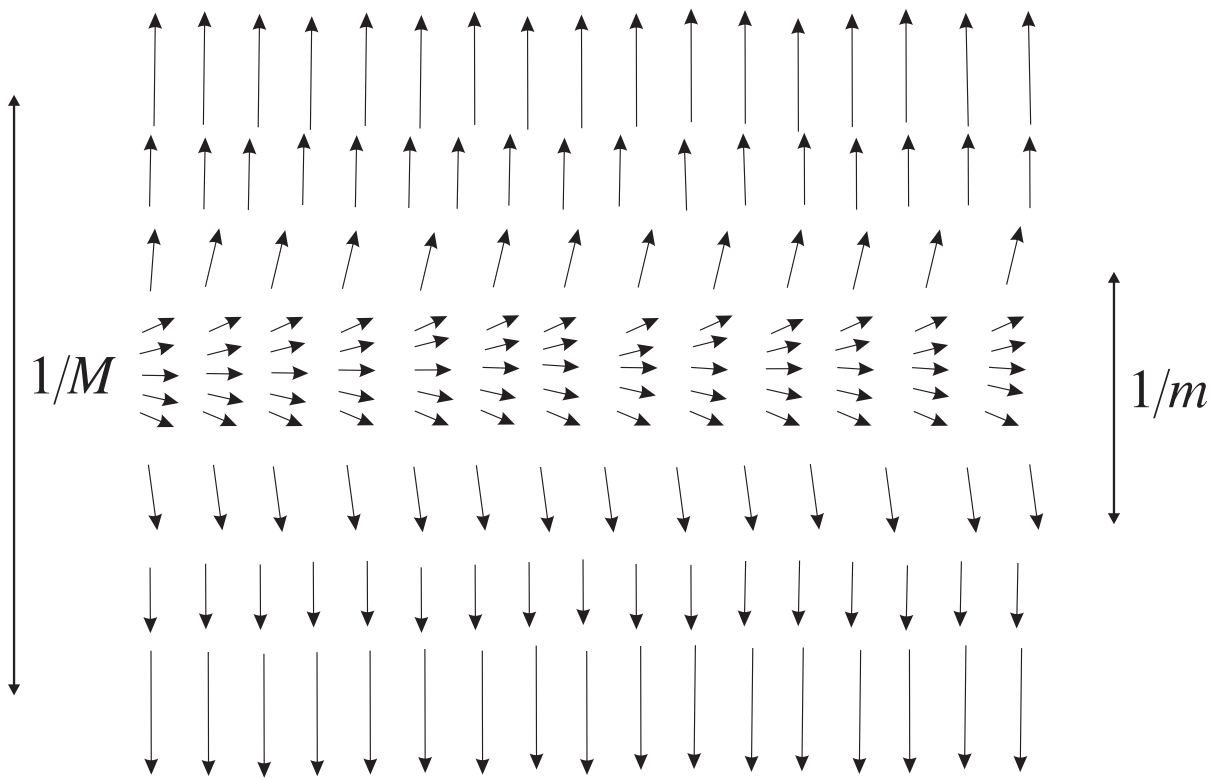}
\caption{The structure of the string (domain wall) in the regime when
the pseudoscalar is heavier than the scalar, $m>M$.}
\label{fig6}
\end{center}
\end{figure}
We will use the same notations, denoting the mass of
the pseudoscalar by $m$ and the mass of the scalar by $M$, but now $m>>M$.
Let us again estimate the string tension and the contributions of the scalar
and a pseudoscalar to it. The width of the region in which the variation of $%
\rho $ takes place is of the order of its inverse mass. The estimate of the
energy density of the $\rho $ field is given by the contribution of the
kinetic term 
\begin{equation}
\sigma _{\rho }\sim M(\mu -\rho _{0})^{2}
\end{equation}
The width of the region in which the phase $\chi $ varies is $\sim 1/m$. In
this narrow strip the radial field $\rho $ is practically constant and is
equal to $\rho _{0}$. The kinetic energy of $\chi $ therefore contributes 
\begin{equation}
\sigma _{\chi }\sim m\rho _{0}^{2}
\end{equation}
Minimizing the sum of the two contributions with respect to $\rho _{0}$ we
find 
\begin{equation}
\rho _{0}\sim {\frac{M}{m}}\mu <<\mu 
\end{equation}
And also 
\begin{eqnarray}
\sigma _{\chi } &\sim &{\frac{M}{m}}M\mu ^{2}  \nonumber \\
\sigma _{F} &=&\sigma _{\rho }\sim M\mu ^{2}
\end{eqnarray}

Now the radial field almost vanishes in the core of the string.
The energy density is contributed entirely by the scalar rather
than by the
pseudoscalar field. Again this is in agreement with decoupling. The heavier
field does not contribute to the energy, even though its values
on the opposite
sides of the wall differ by $O(1)$. Its contribution to the energy
is suppressed by the factor $\rho _{0}^{2}$ which is very small inside
the wall.

We have discussed here the extreme situation $m>>M$. This regime is not
realized in the non Abelian gauge theory. From the lattice simulations we
know that in reality even in the pure Yang Mills case the ratio between the
pseudoscalar and scalar masses is about $1.5$ - not a very large number. The
analysis of the previous paragraph therefore does not reflect the situation
in the strongly coupled regime of the theory. Rather we expect that the
actual profile of the string is somewhere in between Fig.6 and Fig.7
although somewhat closer to Fig.7. The widths of the string in terms of the
scalar and pseudoscalar fields are of the same order, although the scalar
component is somewhat wider. The same goes for the contribution to the
string tension. Both glueballs contribute, with the scalar contribution being
somewhat larger. Of course the spectrum of the 
Yang - Mills theory apart from the scalar
and the pseudoscalar contains many other glueballs. Those
are not included in our effective Lagrangian but their masses are in fact
not that much higher than the masses of the two lowest states. They
therefore also give a nonnegligible contribution to the string
tension, but we have nothing to say about this in the present framework.

Having understood the structure of the domain wall in the two extreme
regimes, we have to ask ourselves what is the interaction between two such
domain walls, or equivalently between two fundamental confining strings. The
answer to this question is straightforward. In the weakly coupled region we
can disregard the variation of $\rho $. For two widely separated strings the
interaction energy comes from the kinetic term of $\chi $. This obviously
leads to repulsion, since for both strings in the interaction region the
derivative of the phase is positive. On the other hand if the pseudoscalar
is very heavy the main interaction at large separation is through the
''exchange'' of the scalar. This interaction is clearly attractive, since if
the strings overlap, the region of space where $\rho $ is different from its
value in the vacuum is reduced relative to the situation when the strings
are far apart.

The situation is therefore very similar to that in superconductivity. The
confining strings in the weakly coupled and strongly coupled regimes behave
like Abrikosov vortices in the superconductor of the second and first kind
respectively. This observation has an immediate implication for the string
tension of the adjoint string. As we have discussed in the previous section
the phase $\chi $ changes from $0$ to $2\pi $ inside the adjoint string. The
adjoint string therefore can be pictured as two fundamental strings running
along each other. In the weak coupling regime the two fundamental strings
repel each other. The two fundamental strings within the adjoint string
therefore will not overlap, and the energy of the adjoint string is twice
the energy of the fundamental one. 
\begin{equation}
\sigma _{Adj}=2\sigma _{F}
\end{equation}

More generally a string connecting two charges of magnitude $n$ (in units of
the fundamental charge) will split into $n$ fundamental strings and its
string tension scales as $n$ 
\begin{equation}
\sigma_n=n\sigma_F  \label{wst}
\end{equation}
This is precisely what was found in \cite{greensite4} in the analysis of the
doubly charged string in the Georgi - Glashow model.

In the strongly coupled region the situation is quite different. The strings
attract. It is clear that the contribution of $\rho $ to the energy will be
minimized if the strings overlap completely. In that case the contribution
of $\rho $ in the fundamental and adjoint strings will be roughly the same.
There will still be repulsion between the pseudoscalar cores of the two
fundamental strings, so presumably the core energy will be doubled. We
therefore have an estimate: 
\begin{equation}
\sigma _{n}=\sigma _{F}+nO({\frac{M}{m}}\sigma _{F}).  \label{sst}
\end{equation}

Again in the pure Yang - Mills limit the situation is more complicated. The
scalar is lighter, therefore the interaction at large distances is
attractive. However the pseudoscalar core size and its contribution to the
tension is not small. In other words $M/m$ is a number of order one. We
expect therefore a dependence on $n$ which is intermediate between eq.({\ref
{wst}) and eq.(\ref{sst})\footnote{%
We would like to mention that some lattice measurements of the string
tension \cite{casimir} suggest the presence of the ''Casimir'' scaling 
\begin{equation}
\sigma _{n}={\frac{n}{2}}({\frac{n}{2}}+1)\sigma _{F}.
\end{equation}
>From the point of view of the effective Lagrangian approach
this dependence is
extremely unnatural. The discrepancy may be due to the fact that the string
tension measurements in 2+1 dimensions are notoriously difficult. This is
due to the difficulty of separating the linear part of the potential from
the Coulomb part which is logarithmic and thus very long range. It is
possible that the Casimir scaling is a property of 3+1 dimensional
theories but 
not the 2+1 dimensional ones. In fact more recent lattice simulations \cite
{faber} in 2+1 dimensions cast doubt on the earlier results in this respect.
It is also possible that even in 3+1 dimensions the Casimir scaling is the
property of the potential in some intermediate region of distances where the
energy is still dominated by the Coulomb potential (which certainly scales
according to the Casimir law) rather than by the string tension.} 
\begin{equation}
\sigma _{n}=f(n)\sigma _{F},\ \ \ \ \ 1<f(n)<n
\end{equation}
Of course this form should be valid only for strings shorter than the
critical length. Adjoint strings must break if they are too long, so for
very large charge separations $f(n)$ vanishes for even $n$ and is equal to $1
$ for odd $n$. In fact it is a nontrivial question whether the critical
length is large enough so that one can sensibly speak about the adjoint
string at all. In the weakly coupled case we saw that in the critical
length is parametrically larger than the string width. The same simple
estimate in the strongly coupled region tells us that the width of the
string is given by the scalar glueball mass, whereas the critical length is
determined by the relation 
\begin{equation}
L_{c}\sigma =2M_{G}
\end{equation}
where $M_{G}$ is the mass of the so called gluelump, or in our language the
core energy of the soliton. One expects that the mass of the gluelump is of
the same order as the glueball mass itself. From the lattice calculations of 
\cite{teper} we learn that $\sigma =.05M^{2}$. So it would appear that 
\begin{equation}
L_{c}\sim 10M^{-1}>>M^{-1}
\end{equation}
and the string should be indeed well identifiable even though there is no
apparent large parameter in the game. 

\section{Discussion.}

In this paper we have analyzed the structure of the confining string as it
emerges from the low energy effective Lagrangian description of 2+1
dimensional $SU(2)$ gauge theory. The mechanism that drives the appearance
of the string is common in the weak and the strong coupling regimes.
Fundamental strings are the domain walls that separate the two degenerated
vacua, as envisioned by t'Hooft. Adjoint strings on the other hand are
somewhat different. They arise due to the fact that the phase of the
disorder parameter has to have a winding number one around a point where the
external adjoint charge is placed. The appearance of the adjoint string is
not tied so closely to the double vacuum degeneracy. Nevertheless in the
weak coupling limit this connection is quite strong. This is due to the fact
that the fundamental strings repel each other and therefore the adjoint
string splits into two fundamental ones running parallel to each other. 

We argued that the strong coupling regime representative of the pure Yang
Mills theory corresponds to a different situation in which the scalar and
the pseudoscalar (the radial part and the phase of the disorder parameter)
have masses of the same order, but the scalar is the lighter one between the
two. In this case the fundamental strings attract. The two fundamental
strings ''forming'' the adjoint string overlap strongly in this case and
practically loose their identity. The qualitative difference between the
weak and strong coupling regimes is therefore similar to the difference
between the superconductors of first and second kind.

Historically there have been two distinct proposals of how to understand
confinement in 2+1 dimensions: t'Hooft's $Z_{N}$ vortex condensation and
Polyakov's screening of the monopole plasma. The former argumentation relies
on the vacuum degeneracy in a crucial way. The latter one does not
particularly care about it. In fact Polyakov's potential for the phase $\chi 
$ in compact QED is $\cos \chi $ rather than $\cos 2\chi $ \cite{polyakov}
and therefore does not even exhibit the $Z_{2}$ symmetry. Still the string
tension is nonzero. Each one of these approaches has its drawbacks.
The t'Hooft mechanism came under criticism because it
can not explain the adjoint string tension. On the other hand Polyakov's
approach suggests that the adjoint string is absolutely stable and does not
break which is obviously incorrect \cite{greensite4}.

The effective Lagrangian approach nicely synthesizes the two. In this
framework it is clear that the vacuum degeneracy is crucial for
understanding the fundamental string. On the other hand the adjoint string
is perfectly happy even if the vacuum state is unique, since it involves an
integer winding of $V$. The adjoint string breaks if it is long enough,
since the theory supports 
dynamical solitons that can screen the external adjoint
charges. In this sense one could perhaps say that the mechanisms for the
fundamental and adjoint string formation are distinct: t'Hooft's mechanism
is relevant for the fundamental string, while Polyakov's mechanism for the
adjoint string. The two are however closely related. Both are represented in
a very simple and intuitive way in the effective Lagrangian through the
properties of the interaction of the disorder field $V$.

Although in this paper we considered only the $SU(2)$ gauge theory, the
discussion can be generalized to any $SU(N)$ group \cite{kovner1}. In this
case the effective theory has a $Z_{N}$ symmetry rather than $Z_{2}$. The
fundamental string is still the domain wall between different vacua whereas
the adjoint string involves an integer winding of the order parameter and
therefore has the same vacuum state on both sides.

One can perhaps use similar ideas to explore confinement in other
gauge theories not necessarily based on the $SU(N)$ group.
The general structure of the ''topological'' mechanism of
confinement as displayed in the effective Lagrangian is the
following. 
Suppose the theory has a
continuos global symmetry $G$  broken spontaneously down to $\ H$.
If  $\pi _{1}(G/H)\neq 1$ the theory has topological solitons which
carry an integer topological charge $Q$. If
in addition $G$ is slightly broken explicitly
these solitons are linearly confined because the topological flux
emanating from a soliton is squeezed into a flux tube by the explicit symmetry
breaking term. This flux tube however breaks by creating a soliton
- antisoliton pair from the vacuum if it is too long. This is the
picture of 
confinement of adjoint charges. The same mechanism confines
fundamental charges if the underlying nonabelian theory has dynamical
fundamental matter.
In a theory without fundamental matter the explicit breaking of $G$ is
not
complete but a discreet subgroup $G^{\prime}$ remains intact.
When $G^{\prime}$ is 
broken spontaneously in the vacuum the set of vacuum states
is discreet. In this case the theory sustains
linelike defects - domain walls between different vacua.
Fundamental charges correspond to externally created
states with 
fractional topological charge
and
serve as termination points for the domain walls. The fundamental
strings are stable since the theory does not have dynamical
fractionally charged solitons.

Finally we want to emphasize the following very important point. The dual
Lagrangian was constructed starting with t'Hoofts idea of the $Z_{N}$
disorder parameter which is a dual variable in the original theory. The
final expression however can equally well (and in fact perhaps more
correctly) be thought of as just a standard low energy effective Lagrangian
of the theory. It is a local Lagrangian written in terms of the lightest
excitation fields, both in the weak and in the strong coupling regimes. As
such it should exhibit all the low energy phenomena, including confinement,
which it does as we have seen. 
In this sense t'Hoofts insight about the $Z_{2}$ dual
symmetry present in the theory should be viewed as a symmetry argument which
constrains the form of the low energy effective Lagrangian, or in a way
determines its universality class.

Recently a similar approach has been suggested in 3+1 dimensional gauge
theory \cite{kovner4}. It was shown that a simple theory of a scalar and a
symmetric tensor field under certain conditions sustains stringlike
solutions which in \cite{kovner4} were interpreted as confining strings.
This theory is naturally viewed as the low energy effective theory of 3+1
dimensional pure Yang Mills theory written in terms of the two lowest
glueball fields, which in this case are a scalar and a symmetric tensor \cite
{glueballs}. It would be extremely interesting to use the analog of t'Hooft
symmetry arguments in 3+1 dimensions to constrain the form of the effective
theory and to see whether the particular conditions required for existence
of the string are imposed on the effective theory by these symmetry
arguments.

{\bf Acknowledgments}

We have benefited from interesting discussions with Luigi Del Debbio, Jeff
Greensite and Mike Teper. A.K. is supported by PPARC Advanced fellowship.
B.R. is grateful to the members of the Oxford Theoretical Physics Department
for hospitality and to Royal Society for financial support during his visit
in April - May 1998 when this work was done.

\end{document}